# Designing with Iontronic Logic Gates - From a Single Polyelectrolyte Diode to Small Scale Integration

Barak Sabbagh, Noa Edri Fraiman, Alex Fish, Gilad Yossifon

**Abstract**

This article presents the implementation of on-chip iontronic circuits via small-scale integration of multiple ionic logic gates made of bi-polar polyelectrolyte diodes. These ionic circuits are analogous to solid-state electronic circuits, with ions as the charge carriers instead of electrons/holes. We experimentally characterize the responses of a single fluidic diode made of a junction of oppositely charged polyelectrolytes (i.e., anion and cation exchange membranes), with a similar underlying mechanism as a solid-state p- and n-type junction. This served to carry out pre-designed logical computations in various architectures by integrating multiple diode-based logic gates, where the electrical signal between the integrated gates was transmitted entirely through ions. The findings shed light on the limitations affecting the number of logic gates that can be integrated, the degradation of the electrical signal, their transient response, and the design rules that can improve the performance of iontronic circuits.

Biological cell membranes contain multiple proteins that act as channels and enable a highly selective exchange of ions and molecules. These channels interact with each other and allow for complex signaling circuits that regulate the transmembrane potential[1]. As compared to artificial circuits, the membrane's threshold response is akin to that of a digital system. Artificial digital logic gates based on ionic signal transmission can thus mimic biological systems, and also function as an electrical circuit. They constitute the emerging research field known as Iontronics[2–5]. Iontronic devices usually contain nanometer-sized fluidic structures (e.g., nanochannels, ion exchange membranes) that exhibit ion permselectivity due to the overlap of their electric double-layers (EDLs)[6]. Their unique electrical behavior enables them to perform in vitro/vivo information processing at the ionic level using iontronic components such as resistors, diodes, capacitors, and transistors[4,7,8]. These nanofluidic components are attracting intensive study, given their fundamental interest and promising applications beyond biomimetic information processing that can contribute to improving chemical and biochemical sensing[9,10] such as for energy harvesting[11], single molecule detection[12], electrokinetic preconcentration[13,14], and brain-machine interfaces[15]. Some of the best-known electrical gate for Boolean logic computation consist of diodes; i.e., diode-based logic gates (DLGs). Diodes are two-terminal components exhibiting a non-linear current-voltage response (I-V) that have a higher conductance in one current flow direction (forward-biasing) than in the reverse direction (reverse-biasing). They can be characterized by their rectification ratio, R, which is defined as the rate between the forward- ($I_F$) and the reverse-biased ($I_R$) currents for opposite voltage polarities ($R=|I_F/I_R|$). Regulation of the electrical current (I) by applying a voltage (V) is critical for the implementation of a logic gate. In DLG, the higher the R, the more precise the control over the current and the higher the gate's performance. Integrating resistors within the circuit provides the constant load resistance needed for the functionality of the gates.

In solid-state electronics, the circuit architecture connecting the diodes and resistors determines the type of DLG. Two types of Boolean functions can be realized by DLG: OR and AND. Each receives a number of logic inputs and returns a single logic output of disjunction/conjunction (for OR/AND DLG, respectively). Specific input and output potential levels are assigned to either 'high' or 'low' binary logic levels that are labeled '1' or '0', respectively. Previous studies have shown that basic ionic circuitry up to a level of sophistication of an individual DLG can be realized with the same principles as solid-state electronics by using either unipolar or bipolar nanofluidic diodes immersed in an electrolyte solution (e.g., aqueous KCl)[16–22]. Unipolar nanofluidic diodes have been made using a geometric symmetry-broken fabricated conical nanopore[23–25] or a funnel-shaped nanochannel[26,27] that slightly favors the ionic current in one direction. However, the maximum R of these diodes rarely exceeds one order of magnitude, which makes it difficult to create an efficient DLG. A more practical realization with a higher R consists of a bi-polar diode made of a junction of two oppositely charged ion permselective regions. Surface functionalized nanochannels[28], field-effect nanochannels[29], nanoparticles[30], and anionic- and cationic exchange membranes (AEM and CEM, respectively) such as polyelectrolytes[16,17,21,31–33] have been used to that end. Under reverse bias, both mobile cations and anions (positivity and negatively charged ions, respectively) are depleted from the junction, which results in significantly decreased conductance. Reversing the direction of the electric field to forward bias transitions the ionic depletion into an ionic enrichment at the junction, which recovers to increased conductance.

The mechanism underlying the nanofluidic bi-polar diode is similar to an electronic solid-state p- and n-type (p-n) junction diode that uses electrons and holes instead of ions as the free charge carriers. Nevertheless, there are fundamental differences between fluidic and solid-state devices in that ion transport is much more complicated. Its complexity stems from electrochemical electron-ion exchanges, the significantly lower mobility of ions compared to electrons, the variety of ionic species, and fluid flow effects[5,34]. All of these have a major impact on DLG performance, and in particular on the integration of several DLGs into multistage circuits, which can further degrade its performance due to current leakage and parasitic resistances. To the best of our knowledge, the well-known behavior of solid-state DLG-based integrated circuits[35] has not been investigated in iontronic DLG-based fluidic circuits.

Here, we report the small-scale integration of a bi-polar polyelectrolyte diode-based iontronic circuit within a microfluidic chip. Whereas the mechanism underlying the operation of a single diode has been studied extensively, we focused on its ability to construct a DLG. The experimentally realized iontronic DLGs were then examined in terms of switching speed, voltage shifting, noise margin, and cascading capabilities, all inspired by the world of solid-state electronics. These enabled us to realize more complex iontronic logic functions with varying circuits by integrating multiple DLGs, where the electrical signal between the integrated DLGs is transmitted entirely through ions.

## Results and Discussion

### Characteristics of a single bi-polar nanofluidic diode

A single bi-polar polyelectrolyte diode interconnected to two microchannels was electrically characterized for 10mM KCl [Fig.1]. At steady-state operation, the transition voltage ($V_{TR}$) closing the open diode occurred at $V_{TR} \approx 0V$ where the ion concentrations at the junction are controlled by the Donnan equilibrium[36]. At positive voltages ($0<V<+1V$), where the diode is considered open, the forward- biased current increased linearly with increasing voltage and reached a value of $|I_F| \approx 1000nA$ at $V=+1V$ [Fig.1B]. At negative voltages ($-1<V<0V$), where the diode is considered close, the reverse-biased current was significantly smaller and almost totally independent of the applied voltage, reaching a value of $|IR| \approx 25nA$ at $V=-1V$. The non-ideal permselectivity of the ion exchange membranes prevented $I_R$ from dropping to zero current under reverse-bias. This non-ideality became worse with higher electrolyte ionic strength, resulting in an enlarged $I_R$ [Fig.S4]. Geometric flaws in the fabricated polyelectrolyte membranes also significantly enlarged the $I_R$ as a result of ion transport that bypassed the membranes [Fig.S5]. The transient I-V scan (100μV/sec) from reverse- to forward- and back to reverse bias ($-1V \rightarrow +1V \rightarrow -1V$) revealed a hysteresis of the current response around 0V due to residual ion enrichment at the junction. This hysteresis grew with increased scanning rate [Fig.S6] while shifting to negative voltages, the transition point at which the diode was closed. This contrasts with solid-state diodes that exhibit a potential barrier (a typical value of $V_{TR} \approx +0.7V$) to open the diode[35]. Increasing the voltage range beyond $|1V|$ introduced additional effects that further complicated the diode's response. Under reverse biasing, water splitting into H+ and OH- generated excess mobile charge carriers that further elevated the $I_R$ once the electric field at the junction exceeded an order of MV/cm[37,38]. Based on the obtained current response, we evaluated the ability to reach these values as crossing a reverse breakout voltage of $V_{BR} \approx -1.4V$. In addition, under forward biasing, generation of ionic depletion regions at the microchannel-polyelectrolyte interfaces due to ion-concentration polarization (ICP) were present as well, resulting in reduced conductance[34]. This led to the appearance of a maximum $I_F$ at a given forward voltage, $V_{MAX}\approx1.2$. Both water splitting and ICP effects resulted in a narrow voltage range of ~|1V| where the rectification ratio reached a maximum value of R=40 at ±1V. Stepwise chronoamperometry (stepping V and monitoring I as a function of time, Fig.1D) was used to measure the current's temporal response at different applied voltages at a switching speed of 3.3mHz. Each step in the potential (step:1,5,6,9,10 V=0V, step:2,4,8,11 V=-1V, step:3,7 V=+1V) simulated a different activation mode that the diode would experience when used to realize a DLG. The findings showed that the current I reached a repeatable equilibrium value for each mode with R=40 at ±1V, regardless of the preceding step that could affect its transient response and RC time constant (defined as the time needed for an RC circuit to reach 63% of its steady-state value). Comparing the time obtained for the diode ($O(10^1 s)$) to that of a typical solid-state diode ($O(10^{-9} s)$)) clearly underscored the difference between ionic mobilities, which were several orders of magnitude smaller than the mobilities of electrons/holes[2]. The diode characteristics that we obtained are summarized in Table 1. When compared to solid-state diode characteristics (depicted in Supplementary Table.S1), it is clear that these emerged as fundamentally different. This highlights the need to further investigate the ability of bipolar polyelectrolyte diodes to realize iontronic DLG as well as integrated circuits made of several DLGs.

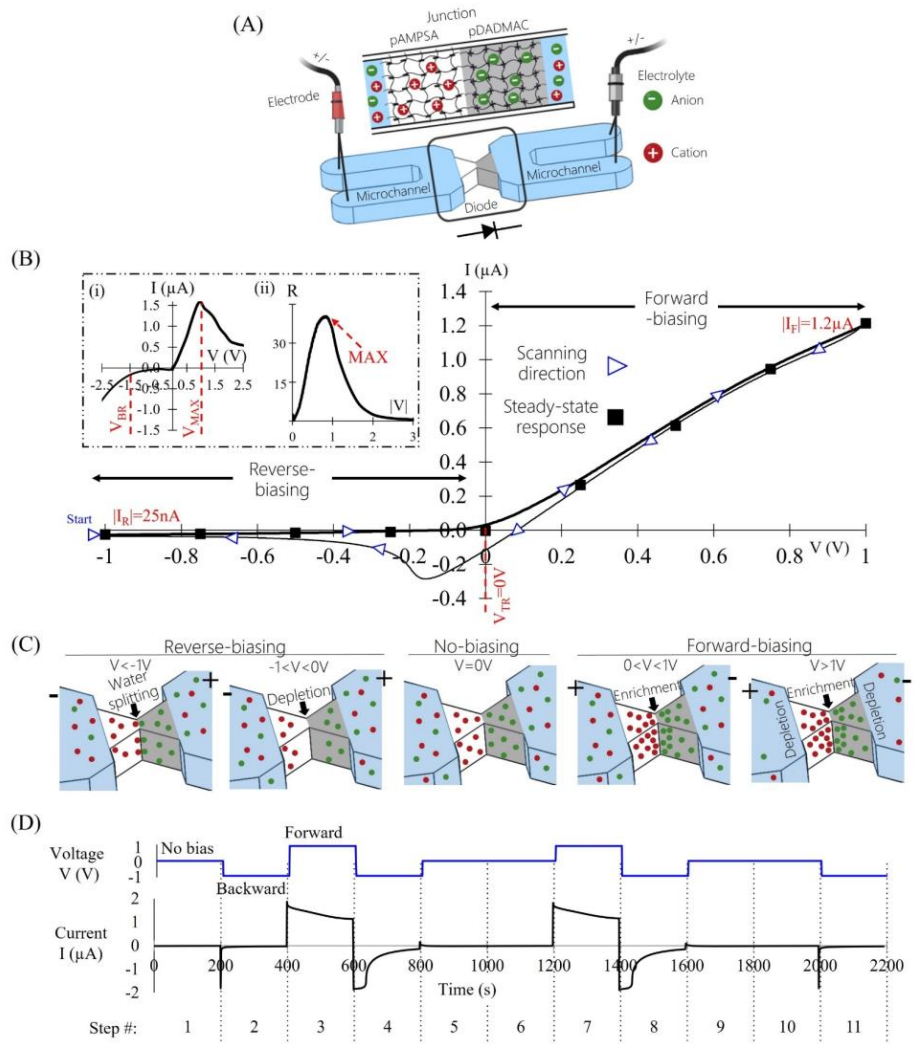

**Figure 1:** Characterization of a single fluidic bi-polar diode's response. **(A)** Schematics of a bi-polar fluidic diode with a cationic-anionic membrane-based junction. The arrow points in the direction of the high conductance current flow of the diode. **(B)** Current-voltage (I-V) response to a voltage scan from -1V to +1V and back to -1V at a scan rate of 100µV/sec. The arrows indicate the scanning direction. Solid black squares indicate the steady-state response of the amperometric measurements under a constant applied V after 500sec. Inset (i): I-V response for -2.5<V< +2.5V. (ii): calculation of the rectification ratio, R, for the V taken from (i), where the maximum R was obtained at ±1V. **(C)** Representative electrolyte ion distribution inside and outside the diode's junction for various Vs. **(D)** Experimental current I response over time (bottom graph, black line) for various voltage V steps (top graph, blue line). Step:1,5,6,9,10 no-bias V=0V, step:2,4,8,11 reverse-bias V=-1V, step:3,7 forward-bias V=+1V.

Besides the bi-polar junction, the interconnecting microchannels have a major impact on the overall response. Ideally, these microchannels should have a negligible effect; however, they exhibited ~80% of the system's total voltage drop under forward-bias and ~2% under reverse-bias (at ±1V) [Fig.S7]. This microchannel-related parasite resistance continued to rise with an increasing number of diodes and the associated number of interconnecting microchannels needed to build the DLGs. The electrical parasite resistance of the microchannel was defined as $\mathscr{R}=L \cdot W^{-1} \cdot H^{-1} \cdot \sigma^{-1}$, where L, W and H were the microchannel length, width, and height, respectively, and σ was the electrolyte conductivity. Since the entire iontronic circuit shared the same electrolyte (i.e., the same conductivity), minimizing $\mathscr{R}$ (without affecting the bi-polar junction) was only possible through local geometry changes of the microchannel. This was achieved by shortening L and enlarging W, while keeping H uniform within the entire microfluidic chip so as not to complicate the fabrication process. In addition to $\mathscr{R}$, the electrodes inserted into the microchannels' inlets may also have added undesired resistance due to their polarization or slow kinetics. To avoid these issues, we used Ag/AgCl electrodes with a large surface area and fast kinetics[39].

| Bi-polar Polyelectrolyte Diode Specifications | | | | @10mM KCl, @ 25°C |
|---|---|---|---|---|
| Characteristic | | Mean Value | Unit | Note |
| Rectification ratio, | R | 28±15 | | at ±1V |
| Reverse current, | $I_R$ | 85±80 | nA | at -1V |
| Reverse breakdown voltage, | $V_{BR}$ | -1.5±0.5 | V | |
| Transition Voltage, | $V_{TR}$ | 0 | V | At steady-state |
| Maximum forward voltage, | $V_{MAX}$ | 1.3±0.5 | V | |
| RC time constant, | $\tau_{RC}$ | 40±38 | s | 66% to steady-state |

**Table 1:** Specification of the bi-polar polyelectrolyte diode. The electrical characterization, including the mean value and its standard deviation, was based on 15 diodes. Typical values for solid-state p-n electronic diodes can be found in supplementary Table S1.

*Characteristics of an individual iontronic DLG (OR/AND)*

Based on the diode's characteristics, we designed and experimentally examined iontronic DLGs [Figure 2]. In our ionic circuits, low (0) and high (1) logical inputs were realized by directly applying 0V(=GND) and +1V(=$V_{DD}$), respectively. A threshold voltage ($V_{TH}$) of +0.5V was defined to distinguish between the two output logic levels. An output potential reading ([Y]) below $V_{TH}$ was defined as a low logical level (0) and above as a high level (1). Unlike standard OR/AND logic gates consisting of two diodes and a resistor, all our circuits were similarly assembled from a symmetric arrangement of three diodes connected by interconnecting microchannels to achieve better performance, where the direction of the diodes dictated the DLGs' functionality. OR DLG was implemented by interfacing the conductive direction of the diodes inwards to the circuit's center, and outwards for AND DLG. Two parallel inputs ([A B]) were introduced into the side diodes (diodes #1,2), while the central diode (#3) was constantly kept closed to act as a load resistance biased to GRD/$V_{DD}$ (for OR/AND DLG, respectively). [Y] was obtained at the center of the circuit between the three diodes. For OR DLG [Figure 2A], if at least one of the inputs was high, the corresponding side diodes became forward-biased. Thus, the current, *I*, passed freely across those diodes with a minimal voltage drop, yielding a high potential at the output measuring point [Y]. [Y] was experimentally measured for various input sequences (a total of $2^2$ possible input sequences), and the stabilized signal after 200s for each input sequence was summarized in a truth table. As expected, low [Y] with a measured voltage approaching GRD (1mV<<$V_{TH}$) was only obtained when both inputs were low ([0 0]). High [Y] was obtained for all other sequences with an average [Y] of 900mV(>$V_{TH}$) with a minimum readout of 877mV ([0 1]). Hence, the applied $V_{DD}$ was degraded across parasite resistances including the forward-biased diodes and microchannels, whereas leakage current through the backward-biased diodes enhanced the degradation to ~ 10% of $V_{DD}$. The small [Y] variations between opposite inputs (e.g. [1 0]:895mV and [0 1]:877mV) were likely due to physical differences between the diodes (see Fig.S8 indicating the variation in R obtained for the different diodes). The symmetric configuration of the DLG minimized these variations by using interconnecting microchannels of the same resistance for any input. For AND DLG [Fig.2B], most of the voltage drop occurred across the reversed-biased side diodes that received low input. The resulting truth table showed output [Y] characterized as a low logic level with voltage variations below $V_{TH}$ between 50 ([0 0]) to 230mV ([1 0]) depending on which and how many inputs were set to low. [Y] only became high with a potential approaching $V_{DD}$ (988mV>>$V_{TH}$) when both inputs were set to high ([1 1]). Examining the transient responses of both OR and AND DLGs revealed similar RC time constants as that of a single diode ($O(10^1 sec)$) due to the gate's parallel inputs. Repeating the same input sequence ([1 0]) at the beginning and the end of the measurement confirmed the repeatability of the DLGs. Plotting all the final output readouts from the individual DLGs revealed the deviations of [Y] from an ideal gate response [Fig.2C]. A deviation from $V_{DD}$ was considered a high voltage range ($VR_H$), and a deviation from GRD as a low voltage range ($VR_L$). In spite of the sufficiently large voltage gap between $VR_H$ and $VR_L$ (~600mV) indicating that the output voltage had attained the correct logical level, the signal deviations limited the number of DLGs that could be cascaded. Investigating the output readouts for a range of input voltages from low to high (0-1000mV) identified the reason why [Fig.2D]: any change in input caused a change in output. High input below 700mV (but above $V_{TH}$) to OR DLG could degrade [Y] below $V_{TH}$ and result in a faulty logic interpretation of a low level. Similarly, a low input above 400mV (but below $V_{TH}$) to AND DLG did not always provide a correct logic level of low [Y]. Hence, this erroneous range (400-700mV) should be avoided as an input, to ensure proper logical functionality of the cascaded DLG. Nevertheless, in cascading DLGs in a series (when each DLG in the series is considered a stage), the output of the preceding DLG drove the input of the subsequent DLG. The fact that the DLGs' acceptable output voltages (0-1000mV) overlapped with this erroneous voltage range would ultimately negate the ability of the subsequent DLG to function as required (defined as a negative noise margin) [Fig.2E].

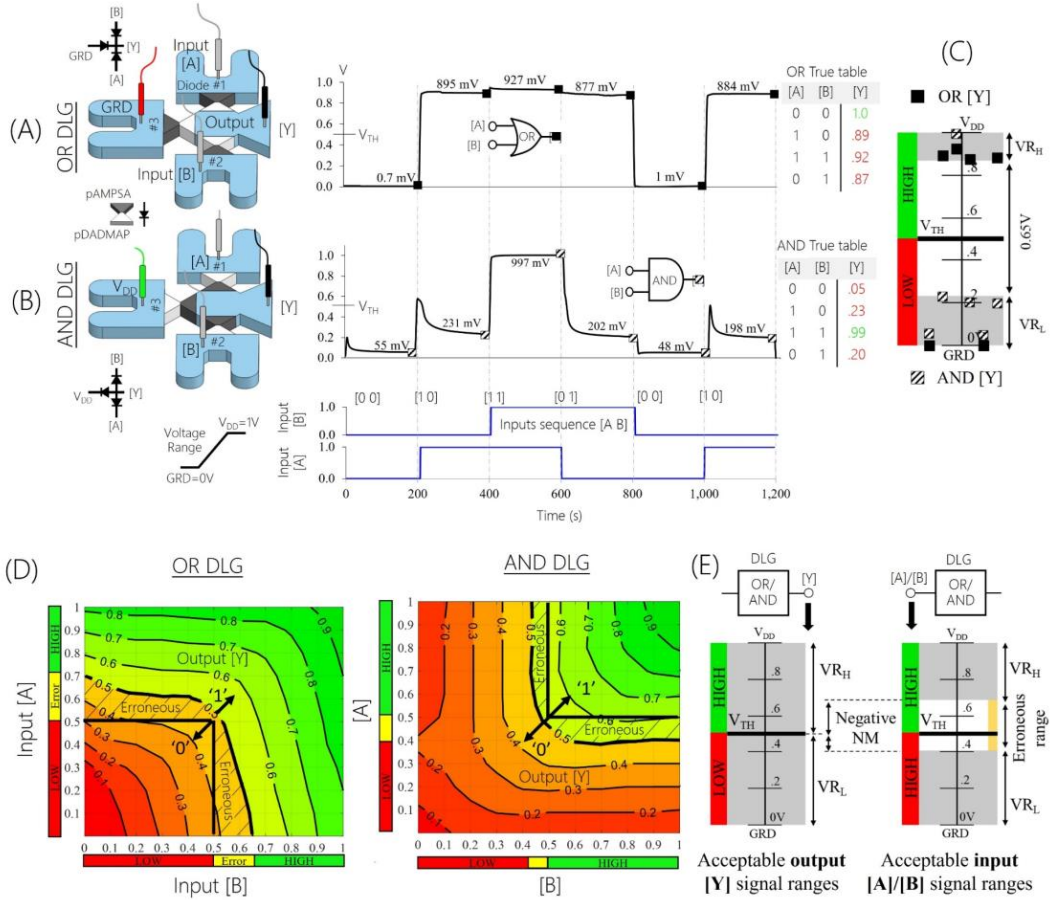

**Figure 2:** Individual DLG (diode-based logic gate) response (OR/AND). **(A)** OR DLG. **(B)** AND DLG. Green and red represent high ('1') and low ('0') logic levels, whereas a threshold voltage of +0.5V(=$V_{TH}$) separates the two. From left to right: Schematics of the fluidic system, containing three bi-polar polyelectrolyte diodes, interconnecting microchannels, two voltage inputs [A] and [B] that receive either 0V(=GRD) or +1V(=$V_{DD}$), and a measured output voltage [Y]. Measured [Y] over time (black line) for various logic input sequences (blue line). Truth table. **(C)** A plot of the output readouts of both DLGs together, where deviations from VDD and GRD (marked in grey) were considered a high voltage range ($VR_H$) and a low voltage range ($VR_L$), respectively. **(D)** Response diagrams of the DLGs for varying voltage inputs (0-1000mV) showing the voltage range of inputs (500-700mV for OR and 400-500mV for AND DLG) that led to an error in the logical interpretation of [Y] (defined as the erroneous range, marked in yellow). **(E)** Noise margin plot of the acceptable output signal ranges versus input signal ranges for both AND and OR DLGs.

*Logical computation by integration of multiple iontronic DLGs*

New logic functions with varying numbers of DLGs, stages, inputs and outputs were realized by integrating multiple DLGs. Each DLG was operated as part of a cascade (integrated) and as an individual DLG (non-integrated by deactivating all other DLGs), allowing us to inspect the effect of integration on the outputs by comparing the output values ($\Delta[Y]=[Y]-[Y]_{Individual}$). Initially, we examined basic integration configurations such as two DLGs connected in parallel (OR||OR) and in series (AND-OR) [Figure 3]. The parallel circuit was designed to consist of a single stage of two OR DLGs with a total of three inputs ([A B C]) and two outputs ([$Y_1$], [$Y_2$]) [Figure 3.A]. A common input ([B]) connected both DLGs by using a common interconnecting microchannel. An effect of the integration of up to +50mV increase of $VR_L$ of both outputs was obtained as inputs ([A C]) drive not only their own DLG through the common interconnecting microchannel. In the series circuit, the output of the $1^{st}$ stage (AND with [$Y_1$]) was connected to one of the two inputs of the $2^{nd}$ stage (OR with [$Y_2$]) [Figure 3.B]. Such an integration resulted in a significant signal deviation of [$Y_2$] expressed in increased $VR_L$ by +250mV. Furthermore, minor variations of [$Y_1$] (±50mV) obtained due to activations of input [C] which associated with input of the $2^{nd}$ stage. Hence, not only the previous stage affected the following one (as was predicted based on the individual DLGs' responses), but also vice-versa. Nevertheless, the results showed that the DLGs could be integrated in both ways while maintaining the correct output logic level for all $2^3$ possible input sequences. None of the output readings fell within the erroneous logical output's voltage range, suggesting that it could be further integrated with additional DLGs. We then successfully realized a more complex circuit that combined integration in series and parallel composed of a two-stage cascade with three DLGs that included an AND DLG that drove two parallelly connected OR DLGs (AND-[OR||OR]) [Figure 3C]. Although a single output ([$Y_1$]) drove two inputs, the deviations remained the same and did not cause a logical failure.

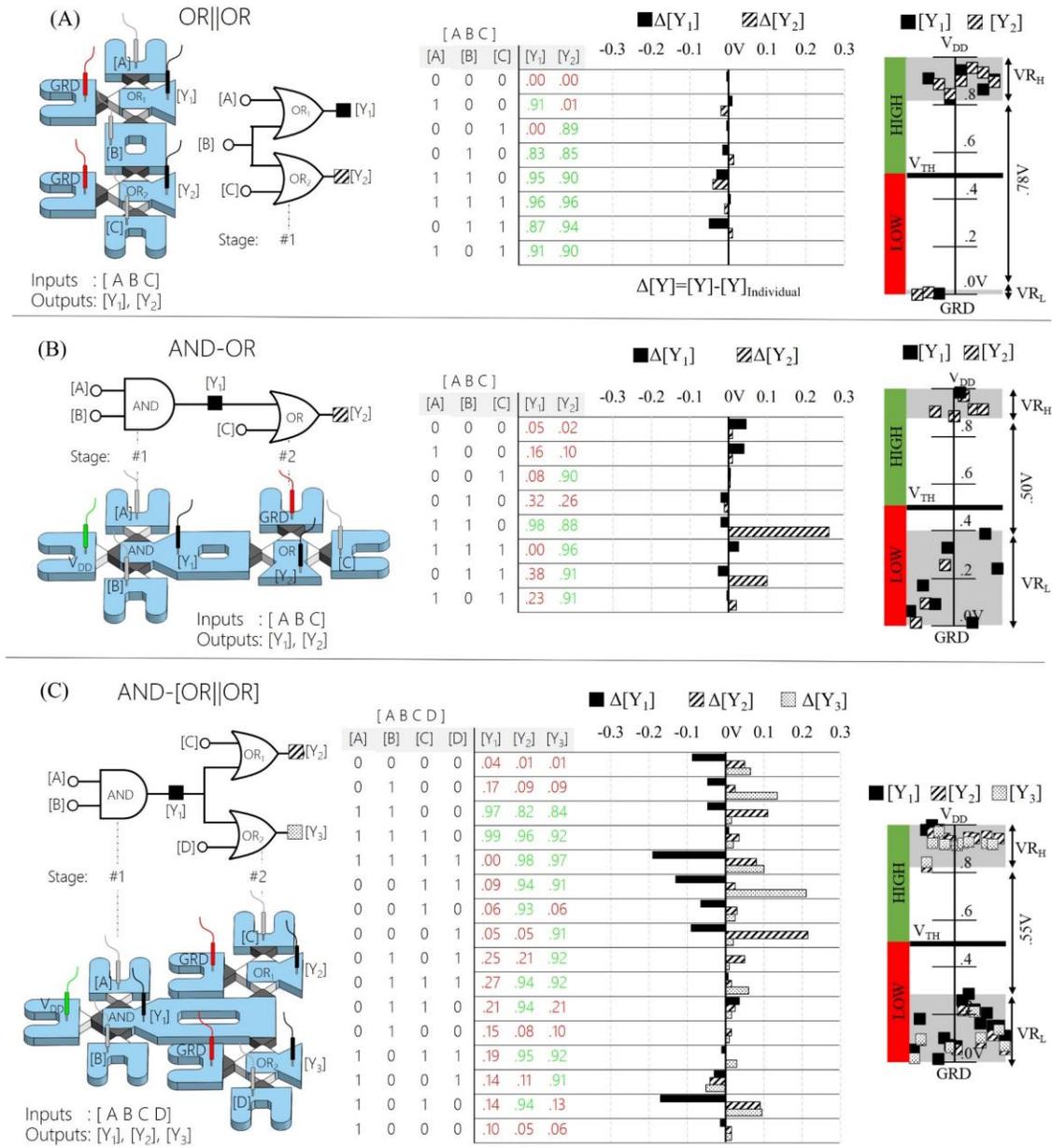

**Figure 3: Integration of multiple diode logic gates (DLGs). (A)** Two OR DLGs connected in parallel with a common input [B] (OR∥OR). **(B)** AND DLG connected in series with OR DLG (AND-OR) where the AND DLG's output [$Y_1$] served as the input of OR DLG. **(C)** A circuit of AND DLG connected in series with two OR DLGs that were connected in parallel (AND-(OR∥OR)). **(A)-(C)** Left to right: Schematic of the fluidic system and logic circuit diagram, truth table, output voltage deviations for each input sequence shown in the truth table, and a plot of the output readouts. Red and green represent low ('0') and high ('1') logic levels, respectively.

The integration limit was reached by cascading four OR DLGs in series (OR-OR-OR-OR, total 4 stages) [Fig.4]. Probing the outputs revealed a monotonic signal degradation at every stage that eventually resulted in a false logic readout [Fig.4A]. Out of 21 examined input sequences, 8 yielded false logic (marked in yellow). For example, degradation from 880mV($>V_{TH}$) at the 1$^{st}$ stage (high [$Y_1$]) to 330mV($<V_{TH}$) at the 4$^{th}$ stage (low [$Y_4$]) although all DLGs should have exhibited a high logic level for the input sequence ([A B C D E]=[1 1 0 0 0]). The false readouts started from the third stage since the second stage's output fell within the erroneous logical output's voltage range (400<[$Y_2$]=600<700mV).Thus, , the behavior of the cascade was predicted based on the individual DLG's response (taken from Fig.2D) [Fig.4B]. Inspecting the same input sequence ([A B C D E]=[1 1 0 0 0]) in the response diagrams, where each diagram represents a stage, revealed a matching trend. We thus used this inspection technique for feasibility studies of other logic functions without having to realize them. For example, a cascade consisting of two OR and AND DLGs connected in series (OR-OR-AND, total of three stages) showed reliable output results [Fig.S9]. For verification, this circuit was experimentally realized and exhibited correct logical readouts as expected. A successful cascading of pairs of AND and OR DLGs with a total number of 5 DLGs integrated in series (AND-OR-AND-OR-AND) was shown to work theoretically based on the response diagram. Potentially this circuit could be extended to include additional DLGs [Fig.S10]. Along with the voltage outputs, the RC time constant was also considered. It was proportional to the number of DLGs through which the ions were transported [Fig.4C]. Although the first stage acquired a saturated signal within tens of seconds ($O(10^1 s)$), the fourth stage only reached saturation after a few hundred seconds $O(10^2 s)$ due to its dependency on the signal of the previous DLGs.

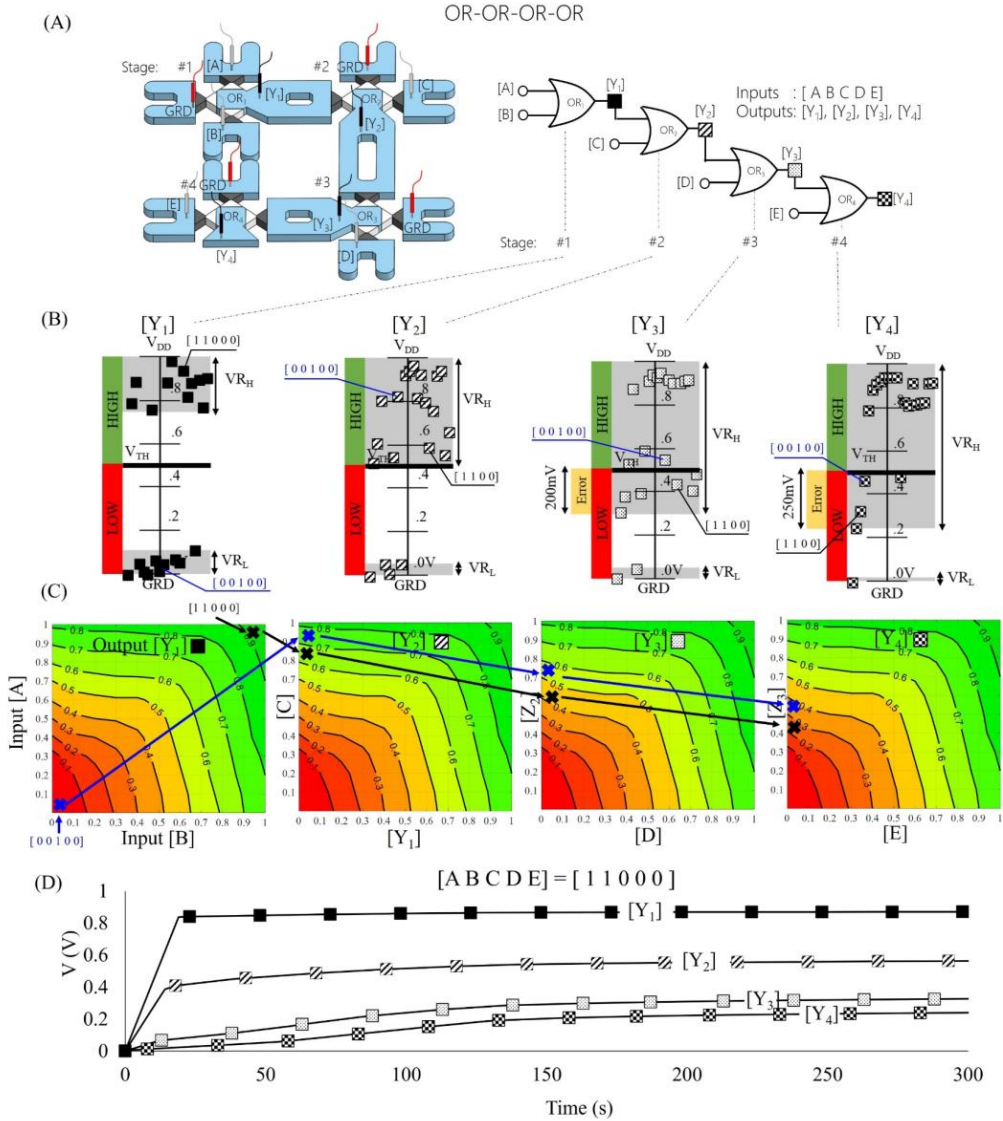

**Figure 4: Integration of four OR gates connected in series. (A)** Gate schematic and its fluidic implementation. **(B)** Plots of the readouts for the output of each stage ([$Y_1$], [$Y_2$], [$Y_3$], [$Y_4$]). Green and red represent logic levels 1 and 0, and yellow indicates a faulty readout. The full truth table can be found in Supplementary Table.S2 **(C)** Prediction of the circuit behavior based on individual DLG responses (response diagrams taken from Fig.2)**.** Two input sequences (of inputs [ A B C D E]) were examined: [ 1 1 0 0 0 ] (black line), and [ 0 0 1 0 0 ] (blue line). X indicates the predicted output voltages, and the arrows indicate the signal propagation path. **(D)** The DLG responses over 300s for high level(=$V_{DD}$) inputs [A] and [B]; the remainder were low level (=GRD) ([ 1 1 0 0 0]).

## Conclusions

We described the implementation of on-chip iontronic circuits using a small-scale integration of diode logic gates (OR/AND gate) consisting of bi-polar polyelectrolyte diodes. The single diode characteristics (e.g., rectification ratio, operation voltages, and RC time constant) determined the operating conditions of the fluidic diode logic gates. Although the mechanism underlying the operation of a bi-polar polyelectrolyte diode and a p-n solid-state diode are similar, their characteristics are fundamentally different. After taking all the circuit components in the micro-chip architecture into account, fluidic diode logic gates that exhibited a consistent and robust differentiation between the logic levels were successfully obtained. We then successfully integrated several such diode logic gates into different circuit architectures, demonstrating for the first time that a real ion-permselective membrane based iontronic integrated circuit can be designed to perform in-chip computation based on various inputs. However, we found that there was a limitation on the number of integrated logic gates, which stemmed from the drift of the output voltage with each subsequent gate due to undesired leakage currents and parasite resistances. Eventually, when the output drift was large enough, it resulted in a faulty logic readout. Thus, simulation tools that take all the integration effects into account, in addition to the responses of a single logic gate, are crucial to the design of an integrated circuit consisting of many logic gates, similar to what is implemented in the design of very-large-scale (VLSI) electronics circuits. Future work could expand the integration capabilities to more complex logic functions with a larger number and types of logic gates consisting of both fluidic transistors and diodes in constructing logic gates that enable amplification of the signal. This type of amplification is essential for signal regeneration throughout the circuit, and can only be achieved by integrating fluidic transistors as well.

## Methods

*Materials.*

3M™ Optically Clear Adhesive 8146-1-ND, 3-(Trimethoxysilyl)propyl methacrylate, 4.2M Diallyldimethylammonium chloride (DADMAC), 4.2M 2-acrylamido- 2-methyl-1-propanesulfonic acid (AMPSA), 2-Hydroxy-4′-(2-hydroxyethoxy)-2-methylpropiophenone, N,N′-Methylenebis(acrylamide), $H_2SO_4(\%):H_2O_2(\%) = 3:1$.

*Fabrication of the microfluidic chip with integrated polyelectrolyte diodes.*

Thin double-sided adhesive tape (approximately 25µm in width) was sandwiched between two glass slides (70x50mm, sigma), and used for patterning the microfluidic channels by cutting the tape (CAMEO silhouette 4). The upper slide was then drilled with holes (1.8mm diameter) as inlets for the microchannels. The polyelectrolyte bi-polar diodes were situated within the microchannels at designated narrow locations (300µm junction minimum width) by polymerization of Diallyldimethylammonium chloride (DADMAC) and (2-acrylamido- 2-methyl-1-propanesulfonic) acid (AMPSA) face to face through UV-light exposure. The interconnecting microchannels were designed such that the solution within could be easily exchanged. Additional information on the fabrication process is provided in Supplementary Figure S1.

*Electrical measurements.*

Silver-silver chloride electrodes (Ag/AgCl, A-M system, 0.015" diameter) were used for the electrical measurements including current-voltage (I-V, scan rate of 100µV/sec) and chronoamperometry, DLG activation by biasing either 0V or +1V to each DLG's input, and open-circuit potential measurements (no net current, I=0) to obtain the DLG's output voltage. A 12-channel automatic relay system (custom-made) was used to expand controllability over two Potentiostats (Keithley 2636, Gamry Reference 3000) towards multiple DLGs. Additional information on the fabrication process and the data acquiring method is provided in Supplementary Figure S2-3.

## Associated content

*Supporting information.*

Fabrication of the microfluidic chip with integrated polyelectrolyte diodes (Fig.S1), acquisition of the diode's output readouts (Fig.S2), the experimental setup (Fig.S3), the effect of electrolyte ionic strength on bi-polar diode performance (Fig.S4), variations in the bi-polar polyelectrolyte diode fabrication and its influence on the rectification factor (Fig.S5), current-voltage response of a single bi-polar diode at two different scan rates (Fig.S6), estimation of the forward- and reverse-biased diode resistance (Fig.S7), rectification ratios of the used bi-polar polyelectrolyte diodes (Fig.S8), successful integration of three diode-logic gates in series (Fig.S9), theoretical output prediction of the integration of five diode-logic gates connected in series (Fig.S10), typical values of solid-state p-n electronic diode (Table.S1), truth table of the circuit in Fig.4 (Table.S2).

## Author information

*Corresponding author.*

**Gilad Yossifon** - School of Mechanical Engineering, Tel-Aviv University, Israel

Email: gyossifon@tauex.tau.ac.il

*Authors.*

**Barak Sabbagh** - Faculty of Mechanical Engineering, Technion–Israel Institute of Technology, Israel

**Noa Edri Fraiman -** Faculty of Engineering, Bar-Ilan University, Israel

**Alex Fish** - Faculty of Engineering, Bar-Ilan University*, Israel*

*Author contributions.*

G.Y., A.F., B.S., and N.E.F conceived the idea of the study. B.S. performed the chip fabrications, experiments, modeling, and data analysis. N.E.F and A.F. assisted in performing the data analysis, and data interpretation in analogy to electronics. B.S and N.E.F contributed equally.  G.Y. supervised execution of experiments, numerical simulations and data analysis. G.Y. and A.F. supervised the planning of the study and the writing of the manuscript. All authors contributed to the preparation of the manuscript.

## Acknowledgments

This work was supported by the Israel Innovation Authority (IIA), Israel Science Foundation (ISF 1934/20). We are grateful to the Technion Russel-Berrie Nanotechnology Institute (RBNI) and the Technion Micro-Nano Fabrication Unit (MNFU) for their technical support. We thank Dr. Baruch Rofman and Mr. Amir Hillman for planning and developing the relay system used in this paper.